\documentclass[submission,copyright,creativecommons]{eptcs}
\usepackage{geometry}  
\usepackage{graphicx}
\usepackage{graphbox}
\usepackage{hyperref}                   % Conflicts with breakurl
\usepackage{amssymb}
\usepackage{epstopdf}
\usepackage{algorithm}
\usepackage{amsmath}
\usepackage{mathtools}
\usepackage[noend]{algpseudocode}
\usepackage{enumitem}
\usepackage{multirow}
\usepackage{hhline}
\usepackage{tabularx}
\usepackage{longtable}
\usepackage{booktabs}
\usepackage{array}
\usepackage{pdflscape}
\usepackage[utf8]{inputenc}
\usepackage[hang,footnotesize,bf]{caption}
\usepackage[acronym]{glossaries} % from sbc paper
\usepackage{cleveref}
\newcommand{\mycomment}[1]{}           % Enable safe multiline comments

\usepackage{listings}
\usepackage{color}

\definecolor{dkgreen}{rgb}{0,0.6,0}
\definecolor{gray}{rgb}{0.5,0.5,0.5}
\definecolor{mauve}{rgb}{0.58,0,0.82}
\definecolor{backcolour}{rgb}{0.95,0.95,0.92}

\lstdefinestyle{mystyle}{
    backgroundcolor=\color{backcolour},
    commentstyle=\color{codegreen},
    keywordstyle=\color{magenta},
    numberstyle=\tiny\color{codegray},
    stringstyle=\color{codepurple},
    basicstyle=\ttfamily\footnotesize,
    breakatwhitespace=false,
    breaklines=true,
    captionpos=b,
    keepspaces=true,
    numbers=left,
    numbersep=5pt,
    showspaces=false,
    showstringspaces=false,
    showtabs=false,
    tabsize=2
}
\usepackage{xargs} 
\usepackage[pdftex,dvipsnames]{xcolor}  % Coloured text etc.
\usepackage[colorinlistoftodos,prependcaption,textsize=tiny]{todonotes}
\newcommandx{\unsure}[2][1=]{\todo[linecolor=red,backgroundcolor=red!25,bordercolor=red,#1]{#2}}
\newcommandx{\change}[2][1=]{\todo[linecolor=blue,backgroundcolor=blue!25,bordercolor=blue,#1]{#2}}
\newcommandx{\info}[2][1=]{\todo[linecolor=OliveGreen,backgroundcolor=OliveGreen!25,bordercolor=OliveGreen,#1]{#2}}
\newcommandx{\improvement}[2][1=]{\todo[linecolor=Plum,backgroundcolor=Plum!25,bordercolor=Plum,#1]{#2}}
\newcommandx{\feedback}[2][1=]{\todo[linecolor=Goldenrod,backgroundcolor=Goldenrod!25,bordercolor=Goldenrod,#1]{#2}}
\newcommandx{\thiswillnotshow}[2][1=]{\todo[disable,#1]{#2}}

\usepackage{tikz}
\usetikzlibrary{trees, arrows.meta, positioning, mindmap, backgrounds, shadows}
\usepackage{forest}

\newacronym{ai}{AI}{Artificial Intelligence}
\newacronym{albert}{ALBERT}{a lite BERT}
\newacronym{api}{API}{application program interface}
\newacronym{apr}{APR}{annual percentage rate}
\newacronym{agi}{AGI}{artificial general intelligence}
\newacronym{bert}{BERT}{bidirectional encoder representations from transformers}
\newacronym{bls}{BLS}{Boneh–Lynn–Shacham}
\newacronym{bn}{BN}{Bayesian network}
\newacronym{cbeth}{cbETH}{Coinbase wrapped staked ETH}
\newacronym{cdf}{CDF}{cumulative distribution function}
\newacronym{cl}{CL}{consensus layer}
\newacronym{clm}{CLM}{causal language modelling}
\newacronym{cpt}{CPT}{conditional probability table}
\newacronym{ctm}{CTM}{Computational Theory of Mind}
\newacronym{dao}{DAO}{decentralised autonomous organization}
\newacronym{dms}{DMS}{data management system}
\newacronym{dos}{DoS}{denial of service}
\newacronym{dvt}{DVT}{distributed validator technology}
\newacronym{eb}{EB}{effective balance}
\newacronym{ef}{EF}{Ethereum Foundation}
\newacronym{eip}{EIP}{Ethereum Improvement Proposal}
\newacronym{el}{EL}{execution layer}
\newacronym{epbs}{ePBS}{enshrined PBS}
\newacronym{ewc}{EWC}{elastic weight consolidation}
\newacronym{ffg}{FFG}{friendly finality gadget}
\newacronym{fxs}{FXS}{Frax share}
\newacronym{ghost}{GHOST}{Greedy Heaviest-Observed Sub-Tree}
\newacronym{gpt}{GPT}{Generative Pre-trained Transformer}
\newacronym{icl}{ICL}{in-context learning}
\newacronym{iot}{IOT}{Internet of Things}
\newacronym{kr}{KR}{Knowledge Representation}
\newacronym{L1}{L1}{Layer 1}
\newacronym{L2}{L2}{Layer 2}
\newacronym{ldo}{LDO}{Lido DAO}
\newacronym{llama}{LLaMA}{Large Language Model Meta AI}
\newacronym{llm}{LLM}{Large Language Model}
\newacronym{lmd}{LMD}{Latest message driven}
\newacronym{lora}{LoRA}{low-rank adoption}
\newacronym{mev}{MEV}{maximal extractable value}
\newacronym{mlm}{MLM}{masked language modelling}
\newacronym{lmm}{LMM}{large multimodal model}
\newacronym{mm}{MM}{MetaMask}
\newacronym{nlp}{NLP}{natural language processing}
\newacronym{nft}{NFT}{non-fungible token}
\newacronym{npt}{NPT}{node probability table}
\newacronym{oecd}{OECD}{Organisation for Economic Co-operation and Development}
\newacronym{opt}{OPT}{open pretrained transformer}
\newacronym{p2p}{PaLM}{Pathways language model}
\newacronym{palm}{p2p}{peer-to-peer}
\newacronym{pbs}{PBS}{proposer builder separation}
\newacronym{pc}{PC}{personal computer}
\newacronym{pdf}{PDF}{probability density function}
\newacronym{peft}{PEFT}{parameter efficient fine-tuning}
\newacronym{pos}{PoS}{proof of stake}
\newacronym{pr}{PR}{pull request}
\newacronym{rag}{RAG}{retrieval augmented generation}
\newacronym{roberta}{RoBERTa}{Robustly optimised BERT approach}
\newacronym{rlhf}{RLHF}{reinforcement learning with human feedback}
\newacronym{rnn}{RNN}{recurrent neural network}
\newacronym{rpl}{RPL}{Rocket Pool}
\newacronym{rtd} {RTD}{replaced token detection}
\newacronym{sft}{SFT}{supervised fine-tuning}
\newacronym{ssf}{SSF}{single-slot finality}
\newacronym{ssl}{SSL}{self-supervised learning}
\newacronym{steth}{stETH}{Lido staked Ether}
\newacronym{tom}{ToM}{Theory of Mind}
\newacronym{ups}{UPS}{uninterruptible power supply}
\newacronym{vae}{VAE}{variational encoder}
\newacronym{vrf}{VRF}{verifiable random function}
\newacronym{web3xai}{Web3xAI}{the intersection of Web3 and AI technologies}
\newacronym{xai}{XAI}{explainable AI}

\title{Intersections of Web3 and AI -- View in 2024
}
\vspace{16pt}
\author{David Hyland-Wood\thanks{Corresponding author}%
	\institute{School of Electrical Engineering and Computer Science, \\ The University of Queensland, Australia}
	\email{david@hyland-wood.org}
\and Sandra Johnson%
	\institute{School of Mathematical Sciences, \\ Queensland University of Technology, Australia}
	\email{sand.johnson@gmail.com}
}
\date{\today}                                         % Activate to display a given date or no date
\begin{document}
\def\titlerunning{Intersections of Web3 and AI - View in 2024}
\def\authorrunning{D. Hyland-Wood \& S. Johnson}
\maketitle

% Abstract 
% -----------
\begin {abstract}
This paper summarises the intersection of Web3 and AI technologies, synergies between these technologies, and gaps that we suggest exist in the conception of the possible integrations of these technologies. The summary is informed by a comprehensive literature review of current academic and industry papers, analyst reports, and Ethereum research community blogposts. We focus our contribution on the perceived gaps and detail some novel approaches that would benefit the blockchain/Web3 ecosystem. We believe that the overview presented in this paper will help guide researchers interested in the intersection of Web3 and AI technologies.
\end{abstract}           

% -------------------------
% Introduction
% -------------------------
\section{Introduction}
\label{sec:introduction}
% ----------------
% Introduction
% ----------------
We first need to address why we are interested in \gls*{ai} at all in the context of Web3 and how general industry trends might affect our industry. Recent advancements in \gls*{ai}, particularly in \glspl*{llm} such as GPT-4, have significant implications for businesses, and especially those in the software development sector. Studies have highlighted a shift in the anticipated impact of \gls*{ai} on labour markets and underscore the increasing relevance of \gls*{ai} in information processing and higher-skilled occupations.

It is important to note that \glspl*{llm} do not constitute the entire field of \gls*{ai}. \gls*{ai} is a complicated and nuanced field with many possible subcategorisations. At a high level, one might consider the subfield of machine learning to include deep neural networks (also called deep learning), supervised and unsupervised learning as well as many other techniques. The subfield of \gls*{nlp} focuses on content extraction, classification, machine translation, question answering and text generation, with \glspl*{llm} being the most modern and successful \gls*{nlp} techniques. Many other subfields exist, including expert systems, machine vision and machine speech systems, planning, and of course robotics. One important reason why it is so hard to categorise \gls*{ai} approaches is the rapidity with which they are combining with each other.

McKinsey \& Company has suggested that \gls*{ai} adoption within corporations should arguably be ``domain focused'' labour \cite{relyea_gen_2024}. That is, a software development organisation should consider the software development domain and look for opportunities to use \gls*{ai} to assist software developers to perform their jobs more effectively.

Importantly to software development organisations, McKinsey also found, ``About 75 percent of the value that generative \gls*{ai} use cases could deliver falls across four areas: Customer operations, marketing and sales, software engineering, and R\&D'' \cite{mckinsey_global_institute_economic_2023}. In other words, business operations and software development activities make up the majority of \gls*{llm} impacts on the labour market, which will naturally affect software development organisations disproportionally. It seems that companies related to science, technology, engineering and mathematics (STEM) that embrace the use of \gls*{ai} are likely to outperform those that do not \cite{maggioncalda_setting_2024}.

The nature of work in the information economy is also anticipated to change drastically. ``Current generative \gls*{ai} and other technologies have the potential to automate work activities that absorb 60 to 70 percent of employees' time today'' \cite{eloundou_gpts_2024}. Thus, we can reasonably expect that jobs in the information economy will look very different in coming decades to what they do today. For software development companies this means a transformation in how software is created, tested, and maintained, with \gls*{ai} taking over routine coding tasks, enhancing productivity, and enabling developers to focus on more complex problem-solving and innovative work.

\subsection{The Impact of \gls*{ai} on the Economy}

How will \gls*{ai} impact the economy and on what timescale? Although we cannot provide complete answers, we can provide some insights and intuitions.

Before \glspl*{llm} came onto the market, the \gls*{oecd} discussed the risks of automation in the context of robotics and the automation of manual tasks \cite{arntz_risk_2016}. They concluded that 9\% of jobs across 21 OECD countries were at risk of automation and emphasised that low-skilled workers were more vulnerable. That is, higher education levels were generally seen as protective against automation. The authors suggested that technological advancements would lead to changes in task structures rather than wholesale job losses, with a need for workforce retraining to mitigate the effects of automation on low-skilled jobs \cite{arntz_risk_2016}.

However, the release of advanced \glspl*{llm} into the market introduced changes due to their natural language capabilities. By 2023, McKinsey Global Institute noted that the scope of \gls*{ai} extended beyond manual tasks to include complex information processing and creative tasks traditionally performed by more highly-educated workers \cite{mckinsey_global_institute_economic_2023}. The report estimated that nearly half of all work activities could be automated by current technologies, affecting jobs across various sectors. Subsequent academic studies have estimated similarly, with \cite{eloundou_gpts_2024} finding 49\% of jobs are likely to be impacted.

McKinsey \cite{mckinsey_global_institute_economic_2023} estimated, ``Generative \gls*{ai} could increase labor productivity by 0.1 percent to 0.6 percent annually over the next 10 to 20 years. When combined with other technologies, automation overall could contribute 0.5 to 3.4 percent annually to productivity growth, assuming labor is redeployed at today's productivity levels and not including general equilibrium effects.'' The impact on productivity would therefore seem to be a key driver of differentiation between those companies that embrace the technology and those who choose not to. Prudent application of these percentages would provide a way to estimate the appropriate level of spending on \gls*{ai} tooling: in order to gain a net value from \gls*{ai} tooling, an organisation needs to spend less on tooling than the productivity gains realised. However, these comparisons are inadequate when \gls*{ai} is further leveraged within an organisation to provide less quantifiable advantages, such as improved customer service, enhanced and novel product development utilising the syngergies of \gls*{ai} and Web3.

Cautions are beginning to emerge from the research community on those promising productivity increases. A recent study by Cui et al noted that Microsoft Copilot was well adopted by both senior and junior developers, but the senior developers showed substantially less benefit from using the product \cite{cui_effects_2024}. The same phenomenon occurred when evaluating seniority by tenure at Microsoft. Worryingly, a 2024 study by Uplevel Labs noted, ``Developers with Copilot access saw a significantly higher bug rate while their issue throughput remained consistent.'' \cite{uplevel_inc_can_2024}. These concerns clearly indicate that the relationship of \glspl*{llm} to productivity is not as clear as initially thought.

To understand the possible time scales for mainstream economic adoption of \gls*{ai} we need to briefly discuss the nature of innovation, how it works and how it unfolds. The biologist Stuart Kauffman developed a ``theory of organization'' to explain biological evolution that he called the ``adjacent possible'' \cite{kauffman_adjacent_1993}. An organism builds itself and interacts with its environment but is limited to what it finds in that environment. That is, what is \textit{possible} is limited by what is \textit{adjacent} to an organism. Further, each new development adds to the possibilities present in the environment. W. Brian Arthur later noted that the same phenomenon occurs in technical innovation \cite{arthur2010nature}. This is important to understand because our current observations of \gls*{llm} orchestration with existing systems is very much an active exploration of the adjacent possible.

Steven Johnson, looking into the nature of innovation, defined the 10/10 rule as ``A decade to build the new platform, and a decade for it to find a mass audience.'' \cite{johnson2011good} He noted that many technologies have followed this pattern, from hardware to software and including the Web. In contrast, he noted that some technologies skip that process and seem to arrive fully formed in very little time. His work explores the differences between those technologies that follow the 10/10 rule and those that manage to proceed faster. The question naturally arises, ``Are \glspl*{llm} a fast or slow technical process?''

One might consider that the original attention mechanism paper was first published in preprint in 2014 \cite{bahdanau_neural_2016}\footnote{First publishing in 2014 and last updated in 2016}. Google's ``Attention is All You Need'' paper followed three years later \cite{vaswani_attention_2023}\footnote{First publishing in 2017 and last updated in 2023}. Here we are in 2024, fully 10 years later. What may seem like a rapid development has been in train for the last decade.

\glspl*{llm} seem, therefore, to be following the 10/10 rule and not the 1/1 step function that has often been inferred in breathless media reporting. We most likely will spend the better part of the next decade absorbing and integrating the capabilities brought by \glspl*{llm}.

\subsection{Unintended Consequences}

There will almost certainly be unintended consequences as \gls*{ai} feeds back on human use of the technologies. For example, human skills in new generations of users may encounter developmental limitations for problems where \glspl*{llm} and other \gls*{ai} forms cannot help. Examples might include new feature development where the desired features have no representation in training sets and/or is not extractable from training. Such eventualities seem historically precedented and  might slow development for certain categories of features and retain the reliance on a few highly-capable individuals who can creatively develop features impossible with \gls*{ai}.

National governments other than those developing new \gls*{ai} systems are also wary of reliance on technology owned, controlled and accessed from countries such as the United States or China. We have seen similar concerns raised regarding reliance on, for example, social media, messaging systems, navigation systems, cloud computing and Internet governance. Smaller countries have become reliant at all levels from households to governments on foreign technical infrastructure. For example, the Reserve Bank of Australia recently noted that ``financial institutions' dependence on `a small number of \gls*{ai}' could `create vulnerabilities due to a single point of failure'' ' \cite{sadrolodabaee_china_2024}.

\subsection{Structure of this Paper}

We begin this paper by providing a comprehensive literature review covering the intersection between Web3 and \gls*{ai} technologies in section~\ref{sec:literaturereview},~\nameref{sec:literaturereview}. We focus especially on how these technologies could complement each other.

We discuss the landscape of \gls*{web3xai} and identify the key thought leaders. We follow by identifying the key interlinking technologies, and cover various proposals for the application of \gls*{ai} to Web3. We then categorise the various proposals for \gls*{web3xai} integration:

\begin{itemize}
\item Sustainability
\item Non-\gls*{llm}-based \gls*{ai} with Web3 utility
\item Ownership and copyright applications
\item Philosophical and ethical implications
\item Future prospects and research directions identified by others
\end{itemize}

In section \ref{gapanalysis}, \nameref{gapanalysis}, we identify new potential research directions not found in the literature review. In section~\ref{risksandmitigations},~\nameref{risksandmitigations} we identify a series of common risks when working with \gls*{ai} in general and \glspl*{llm} in particular. In section~\ref{researchdirections},~\nameref{researchdirections}, we suggest research directions based on our gap analysis and lessons learned so far. Finally, we conclude by summarising the key points discovered to date.

% --------------------------------
% Literature Review
% --------------------------------
\section{Literature Review}
\label{sec:literaturereview}

% ------------------------------------
% Overview
% ------------------------------------
% --------------------------
\subsection{Overview}
% --------------------------
\label{sec:litrevoverview}

The proposed integration of blockchain and artificial intelligence (AI) represents a potential transformative shift across various industries, offering prospects for enhanced security, transparency, and efficiency. This literature review examines the current state of research on the proposed convergence of Web3/blockchain and AI technologies, focusing on their potential applications, benefits, challenges, and future prospects \cite{palaiokrassas_machine_2023,achuthan_sustainable_2024,baioumy_ai_2024,bhumichai_convergence_2024,marengo_future_2024}

Baioumy \& Cheema suggest that we can view the crossover between Web3 and AI as falling into one of three categories \cite{baioumy_ai_2024}:

\begin{enumerate}
\item Blockchain technology benefiting AI technology (AI expands its capabilities)
\item AI technology benefiting Blockchain technology (Blockchain expands its capabilities)
\item AI \& Blockchain technologies co-existing without constructive interaction (both serve a useful purpose, but act independently)
\end{enumerate}

Vitalik Buterin has also waded in on the synergies of these two important technologies in his blog post, \textit{The promise and challenges of crypto + AI applications} \cite{buterin_promise_2024}. He categorised interactions between AI and blockchains as follows (quoting from his blog post):

\begin{enumerate}
\item AI as a player in a game [highest viability]: AIs participating in mechanisms where the ultimate source of the incentives comes from a protocol with human inputs.
\item AI as an interface to the game [high potential, but with risks]: AIs helping users to understand the crypto world around them, and to ensure that their behavior (ie. signed messages and transactions) matches their intentions and they do not get tricked or scammed.
\item AI as the rules of the game [tread very carefully]: blockchains, DAOs and similar mechanisms directly calling into AIs. Think e.g. "AI judges"
\item AI as the objective of the game [longer-term but intriguing]: designing blockchains, DAOs and similar mechanisms with the goal of constructing and maintaining an AI that could be used for other purposes, using the crypto bits either to better incentivise training or to prevent the AI from leaking private data or being misused.
\end{enumerate}

Yunlong and Jie discuss the fact that AI, game theory and blockchain have the ``potential to create an economically sustainable cloud ecosystem'' \cite{yunlong_incentive_2024}. Others that attempted to define the intersection of Web3 and AI included \cite{ressi_ai-enhanced_2024,perrault_artificial_2024}.     

% ------------------------------------
% Interlinking technologies
% ------------------------------------
% -----------------------------------------------
\subsection{Interlinking Technologies}
% -----------------------------------------------
\label{sec:litrevinterlinking}

\href{https://www.turing.com/}{Turing Enterprises, Inc.} identifies the two technologies interlinking in the following ways \cite{turingcom_future_2024}:
	
\begin{enumerate}
\item Transparent data source annotations for AI training data
\item Management of unsupervised AI training and operations in a distributed manner
\item Development of privacy-protecting AI systems, e.g. via cryptographic techniques such as zero knowledge proofs
\item Blockchain technology could assist with obtaining sufficient computing power and storage to run AI applications
\item Improving the security of smart contracts via anomaly detection
\item Improvements in reading efficiency that is often sacrificed by a write-intensive \gls*{dms} such as a blockchain.
\item Visibility into the authenticity and provenance of AI training data (``explainable AI'')
\item Augmentation of humans ability to read and summarize large volumes of information
\item Automation of business processes that span different parties, e.g., to resolve disputes and select the most sustainable shipping methods.
\end{enumerate}

Although item 4 may be up for some interpretation, we suggest that Turing's intention was for blockchains to assist economically with the procurement of computing resources via on-chain or L2-based searching (e.g., as Filecoin does for storage) followed by on-chain bidding and payments for such resources via smart contracts.

% --------------------------------------
% Applications of Web3 and AI
% --------------------------------------
\subsection{Applications of Web3 and AI}
% ---------------------------------------
\label{sec:litrevapplications}

Those proposing various domain-specific applications of Web3 and \gls*{ai} interactions \cite{bhumichai_convergence_2024,bentley_combination_2024,nguyen_generative_2024,buterin_promise_2024,ressi_ai-enhanced_2024,yunlong_incentive_2024,zhao_approach_2023} illustrate the wide interest in a Web3 and AI convergence.

Of particular note are those who have proposed enhancements to cybersecurity systems on-chain with the application of \gls*{ai} \cite{achuthan_sustainable_2024,fernandez-becerra_enhancing_2024} including those with a focus on cybersecurity education \cite{ali_educational_2024} and payment security \cite{agrawal_harnessing_2024,josyula_role_2021}. Attempts to identify potential fraudulent transactions in blockchain data \cite{mbula_mboma_assessing_2023,elmougy_anomaly_2021,farren_low_2016} extend the realm of payment security validation to the blockchain (\gls*{L1}) layer by looking across many transactions.

The use of new \gls*{ai} techniques to improve the state of software auditing \cite{david_you_2023,du_evaluation_2024,ma_combining_2024}, although not ready to stand on their own, may provide valuable adjuncts to existing auditing techniques.

At least one analyst suggests that, ``onchain applications and protocols will likely find a new class of onchain participants via AI agents'' and reports that several organisations are actively working towards that goal \cite{bloomberg_dissecting_2024}.

\begin{figure}[htbp]
\begin{center}
\includegraphics[width=0.6\linewidth]{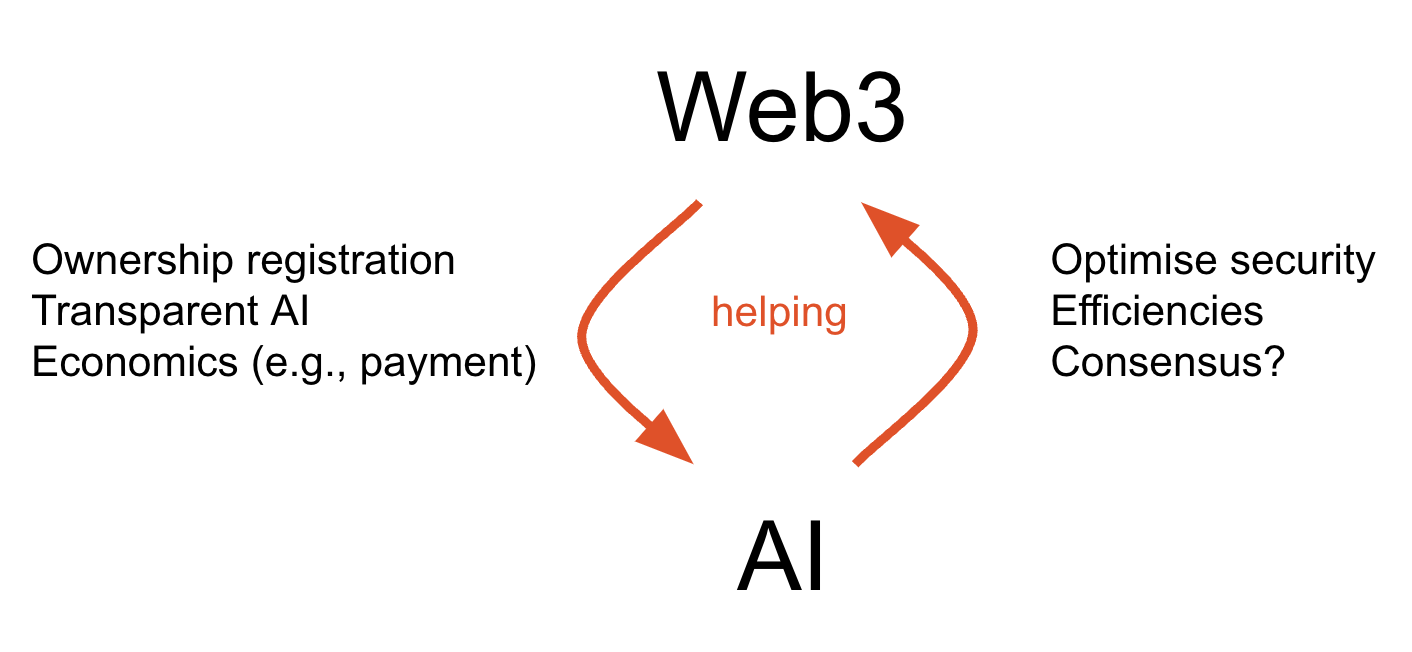}
\caption{Evolving relationship between Web3 and AI}
\label{fig:relationship}
\end{center}
\end{figure}

We summarise the major directions of the evolving relationship between Web3 and \gls*{ai} in Figure \ref{fig:relationship}.

% ---------------------------------------

% ---------------------------------------
% Sustainability
% ---------------------------------------
% ---------------------------------------
\subsection{Sustainability}
% ---------------------------------------
\label{sec:litrevsustainability}

It is no secret that blockchains and modern AI systems consume large amounts of electrical power. Using AI technology itself to address this concern by searching for more efficient blockchain techniques has been widely proposed \cite{achuthan_sustainable_2024,bentley_combination_2024,bhumichai_convergence_2024,chaudhari_nft_2023}.

% ---------------------------------------
% Non-LLM-based AI
% ---------------------------------------
% ---------------------------------------
\subsection{Non-LLM-based AI}
% ---------------------------------------
\label{sec:litrevnonllm}

\begin{figure}[htbp]
\begin{center}
\includegraphics[width=0.8\linewidth]{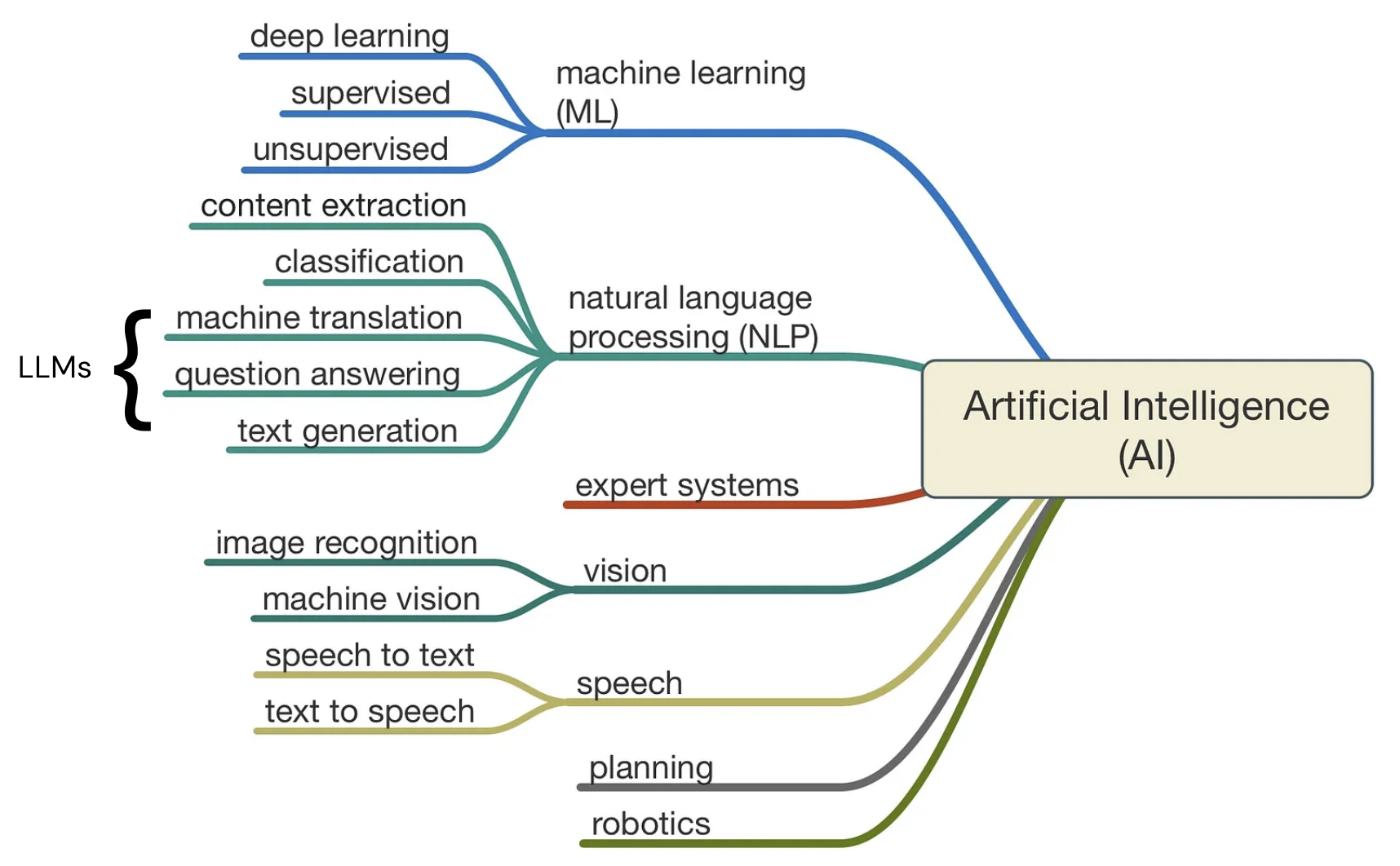}
\caption{There are many branches of \gls*{ai} (After: Chethan Kumar GN, Incedo Technologies)}
\label{fig:ailandscape}
\end{center}
\end{figure}

\gls*{ai} is not synonymous with \glspl*{llm}. The field of \gls*{ai} is broad and in many areas, relatively mature. Figure \ref{fig:ailandscape} imperfectly summarises the major subfields of \gls*{ai}. In fact, 

A number of researchers have been applying traditional machine learning techniques to blockchains \cite{elmougy_anomaly_2021,farren_low_2016,xu_vasa-1_2024,palaiokrassas_machine_2023} and considering the impact of both blockchains and broader AI techniques to the field of \gls*{iot}.

Another example of older approaches to \gls*{ai} is knowledge representation \gls*{kr}. \gls*{kr} may be combined with \glspl*{llm} to provide knowledge of structured data such as vocabularies of specified terms, thus creating formalised entities that an \gls*{llm} can act upon. \cite{schoch_nl2ibe_2024}

% ---------------------------------------
% Ownership and Copyright
% ---------------------------------------
% ---------------------------------------
\subsection{Ownership and Copyright}
% ---------------------------------------
\label{sec:litrevownership}

Blockchains, with their permanent record of transactions, have long been proposed as repositories of ownership information. In this context, Potluri et al \cite{potluri_securing_2023} and Mourya \cite{mourya_strengthening_2023} specifically address blockchain-backed copyright registration for arbitrary intellectual property, including AI-generated content.

An obvious implementation approach for ownership registration on blockchains is to extend the existing usage of \gls*{nft} to arbitrary intellectual property. Such an approach has been proposed or acted on by many \cite{darwish_nfts_2024,chaudhari_nft_2023,jia_nfts_2023,ferro_digital_2023,lee_role_2024,umiyati_legal_2023}.

% ----------------------------------------------------------------
% Philosophical and Ethical Implications
% ----------------------------------------------------------------
% ----------------------------------------------------------------
\subsection{Philosophical and Ethical Implications}
% ----------------------------------------------------------------
\label{sec:litrevphilosophical}

The ability of some \glspl*{llm} to pass \gls*{tom} \cite{strachan_testing_2024, kosinski_evaluating_2024} and Turing Tests \cite{biever_chatgpt_2023, mei_turing_2024} suggests support for the \gls*{ctm}, that cognition may be substrate independent. These findings challenge biological essentialism and open new avenues for creating sophisticated AI systems capable of human-like reasoning and interaction. Viewed another way, these studies could be taken to provide evidence for those critical of both the Turing Test and Theory of Mind tests in assessing cognition in humans and animals.

However, it should be noted that \glspl*{llm} by themselves have no self monitoring (also called phenomenal consciousness or subjective experience) or internal, updatable model of their external environment (that is, a model of itself as a being in a world). Both of these conditions are required in some reasonable theories consciousness \cite{kuhn_landscape_2024}. Those limitations alone may be regarded as evidence that \glspl*{llm} lack any form of consciousness as that term is currently understood.

Language models pre-trained on large text corpora that are highly likely to contain toxic and inappropriate content, are known to pass these biases on, or worse amplify them, when generating query responses and text \cite{touvron2023llamaopenefficientfoundation}. The bias can present itself in various forms such as discrimination based on race, gender, disability, nationality or religion \cite{nangia-etal-2020-crows, touvron2023llamaopenefficientfoundation}. To this end researchers developed a challenge dataset,  CrowS-Pairs, that was crowd sourced using Amazon Mechanical Turk (MTurk), to measure the extent of bias in \gls*{mlm} \cite{nangia-etal-2020-crows}.          

% ----------------------------------------------------------------------
% Future Prospects and Research Directions
% ----------------------------------------------------------------------
% ----------------------------------------------------------------------
\subsection{Future Prospects and Research Directions}
% ----------------------------------------------------------------------
\label{sec:litrevfuture}

Major themes suggested for future research focus on interdisciplinary collaboration, regulatory compliance and ethical considerations \cite{ali_educational_2024,marengo_future_2024,jia_nfts_2023,umiyati_legal_2023,ferro_digital_2023,josyula_role_2021} and future economic impacts, including mechanisms to record and validate ownership \cite{yunlong_incentive_2024,josyula_role_2021,umiyati_legal_2023,lee_role_2024,ferro_digital_2023,chaudhari_nft_2023,potluri_securing_2023,darwish_nfts_2024,mourya_strengthening_2023}.

Some researchers proposed what Buterin \cite{buterin_promise_2024} called ``singleton'' integrations \cite{nguyen_generative_2024,ressi_ai-enhanced_2024} or ``a single decentralized trusted AI that some application would rely on for some purpose''. Buterin warned that such approaches were ``the most challenging to get right''.

There was substantial interest in improving AI's capabilities to improve auditing of software systems for vulnerabilities \cite{elmougy_anomaly_2021,farren_low_2016,david_you_2023,ma_combining_2024,mbula_mboma_assessing_2023}.

Several researchers called for the creation of ``transparent'' AI systems where the training data used was serialised on blockchains \cite{baioumy_ai_2024,fernandez-becerra_enhancing_2024,marengo_future_2024} and blockchain-backed marketplaces of \gls*{ai} systems \cite{baioumy_ai_2024,bentley_combination_2024}.

At least one researcher remained concerned that we are not yet exploring all possible themes for Web3 and \gls*{ai} integrations \cite{achuthan_sustainable_2024}, a sentiment shared by the authors of this paper.

Additional research was proposed in enhancing the abilities of AI systems for:

\begin{itemize}
\item \gls*{dao} governance via \gls*{ai} \cite{baioumy_ai_2024}
\item investigating blockchain consensus algorithms using \gls*{ai} \cite{bentley_combination_2024}
\item enhancing blockchain security via quantum cryptography \cite{agrawal_harnessing_2024}
\item federated \gls*{ai} learning \cite{baioumy_ai_2024,bentley_combination_2024}
\item scalability \cite{bhumichai_convergence_2024}
\end{itemize}

% ----------------------------------------
% Research Gap Analysis
% ----------------------------------------
\section{Research Gap Analysis}
% -------------------------------
% Research Gap Analysis
% -------------------------------
\label{gapanalysis}

As discussed in Section \ref{sec:litrevoverview} \nameref{sec:litrevoverview} in the \nameref{sec:literaturereview}, Buterin conceived of four categories and Baioumy of three categories of \gls*{web3xai} integration. We can add at least one additional one: \gls*{ai} as a social disrupter where Web3 can help reduce negative consequences. Some examples would include on-chain attributions of ownership of intellectual property such as videos, news stories, contracts, photos coupled with automated use of such checks (e.g., for fact-checking or attribution confirmation). Turing Enterprises \cite{turingcom_future_2024} and Potluri \cite{potluri_securing_2023} both suggest such an approach but have yet to explore it. This is a clear area for additional research beyond mere attribution.

We note that the concept of ``Singleton'' Webx\gls*{ai} integration is poorly defined despite some initial proposals by Ressi et al \cite{ressi_ai-enhanced_2024} and Nguyen et al \cite{nguyen_generative_2024}. No doubt many more proposals will follow in due course. The concept seems likely to mature and implementations are likely to follow whether or not they ever reach the ``single decentralized trusted AI that some application would rely on for some purpose'' envisioned (and warned against) by Buterin \cite{buterin_promise_2024}. There is simply a lot of room between the current state of the art and such a sweeping vision.

There generally seems to be a surprising gap in published research at the L2 level and yet L2s are arguably the easiest place to explore \gls*{ai} integration. Certainly algorithmic trading, compliance and usage assistants may touch the L2 level but how might new business models emerge from \gls*{web3xai} integration at the L2 level? Only time will tell.

% ------------------------------------
% Risks and Mitigations
% ------------------------------------
\section{Risks and Mitigations}
\label{risksandmitigations}
% ----------------------------------------------
% Organisational Risks
% ----------------------------------------------
% ----------------------------------------------
\subsection{Organisational Risks}
% ----------------------------------------------
\label{organisationalrisks}

It would be na\"{i}ve to think that the integration of \gls*{ai} systems into organisations' existing systems and processes is going to go smoothly. Consider as a salient example a recent newspaper report suggesting that pilot implementations of Microsoft Copilot have been stopped in a number of large companies due to data governance issues \cite{claburn_top_2024}. The particular issues described related to inappropriate access to proprietary information by both the \glspl*{llm} and subsequently by employees. This should not be a surprise given that previous research has demonstrated that, ``Mobilizing an organization to adopt data governance has proven challenging in practice'' \cite{benfeldt_data_2020}. In fact, data governance has been a persistent challenge for organisations large \cite{ladley_data_2019} and small \cite{begg_exploring_2012}.

Organisations should consider creating a risk framework for their specific \gls*{ai} projects and integrations. Slattery et al have assisted such an effort by collecting more than 700 risks extracted from 43 existing frameworks and creating two taxonomies of \gls*{ai} risks \cite{slattery_ai_2024}. Pulling relevant risks from this near-comprehensive collection of risks should ease the burden of creating new risk frameworks and help reduce the number of missed or unconsidered risks.

% -----------------------------------------------------
% Explainable AI
% -----------------------------------------------------
% ----------------------------------------------
\subsection{Explainable AI (XAI)}
% ----------------------------------------------
\label{sec:explainableAI}
Accompanying the proliferation and complexity of \gls*{ai} tools and models, is an increased perception that \gls*{ai} models are in effect ``black boxes'', ultimately undermining trust in the validity and authenticity of their output and decision making and hence an inherent resistance to apply this technology to crucial but sensitive areas such as healthcare and defence, despite its potential to be immensely valuable in these domains \cite{gunning2019, rai_explainable_2020,linardatos_pantelis_explainable_2021, fernandez-becerra_enhancing_2024}. This includes the requirement for transparency of the origin and content of pre-training data to alleviate substantiated concerns over potential biases inherent in publicly available datasets \cite{baioumy_ai_2024,fernandez-becerra_enhancing_2024,marengo_future_2024}.

To address the general lack of trust, accountability, and transparency coupled with the inability to humanly comprehend model reasoning, a research field known as explainable AI (\gls*{xai}) has emerged. \gls*{xai} encompasses methods and techniques to make \gls*{ai} model decisions and actions understandable to humans \cite{linardatos_pantelis_explainable_2021,yang_survey_2023}. Moreover, the ability to communicate their rationale to a human, while also highlighting their own strengths and weaknesses are central to engendering trust in \gls*{ai} methodology \cite{gunning2019}. 

\begin{figure}[htbp]
\begin{center}
\includegraphics[width=0.7\linewidth]{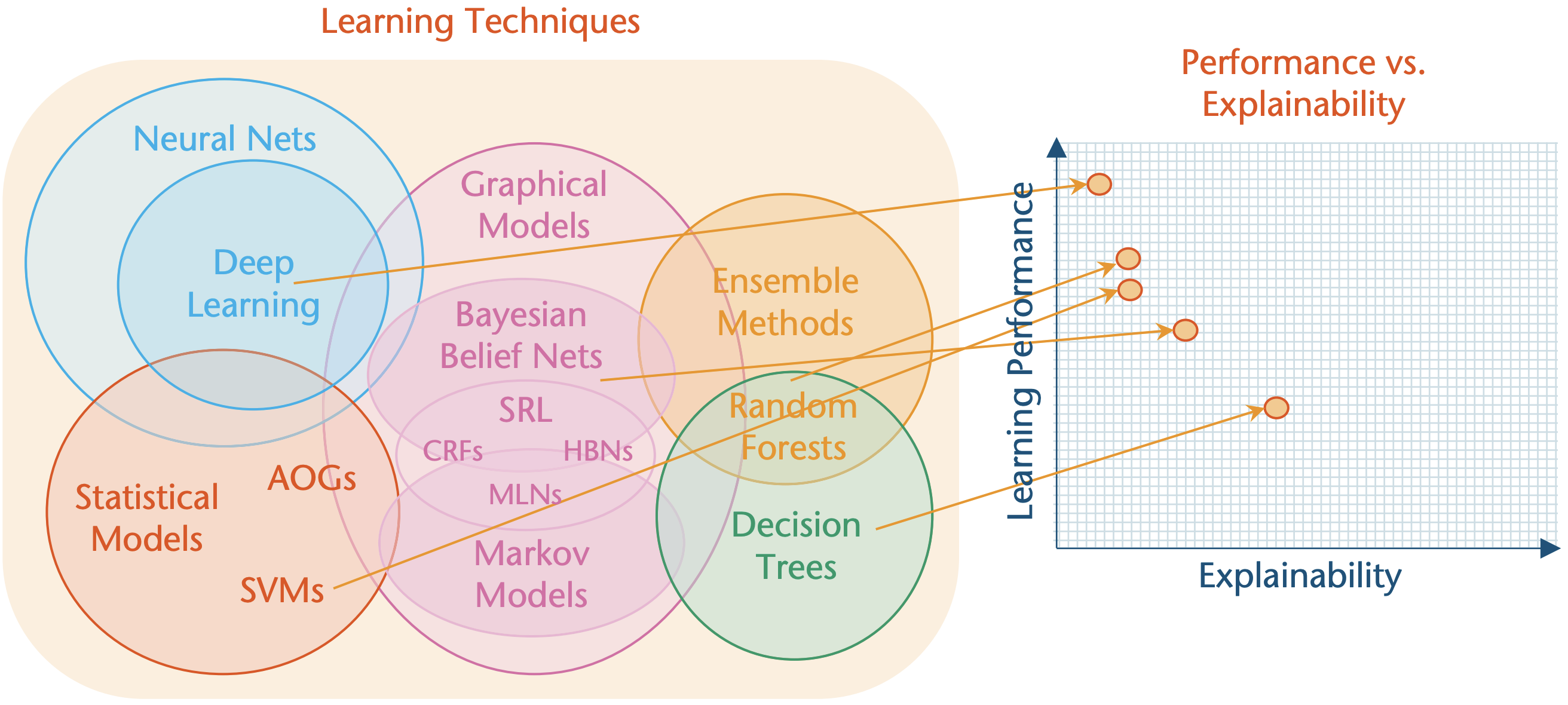}
\caption{Model learning performance vs model explainability (image by \cite{gunning2019})}
\label{fig:learningvsexplainability}
\end{center}
\end{figure}

There is a delicate balance between machine learning and statistical model learning performance and the ease of interpretability of their predictions. Without any complementary \gls*{xai} techniques, Figure~\ref{fig:learningvsexplainability}, page~\pageref{fig:learningvsexplainability}, visually represents the learning performance and explainability of several groups of models. Although machine learning models like deep learning neural networks are the most performant at learning, they are unfortunately also the most obtuse to users. Ensemble methods such as random forests and statistical models such as state vector machines (SVMs) have lower learning performance but slightly better explainability. Probabilistic graphical models such as Bayesian networks are more humanly understandable, but it comes at a slight degradation in learning performance. Statistical models such as linear regression and decision trees are least performant in terms of learning but are readily comprehensible by humans \cite{gunning2019,linardatos_pantelis_explainable_2021} (Figure~\ref{fig:learningvsexplainability}, page~\pageref{fig:learningvsexplainability}).

\begin{figure}[htbp]
\begin{center}
\includegraphics[width=0.65\linewidth]{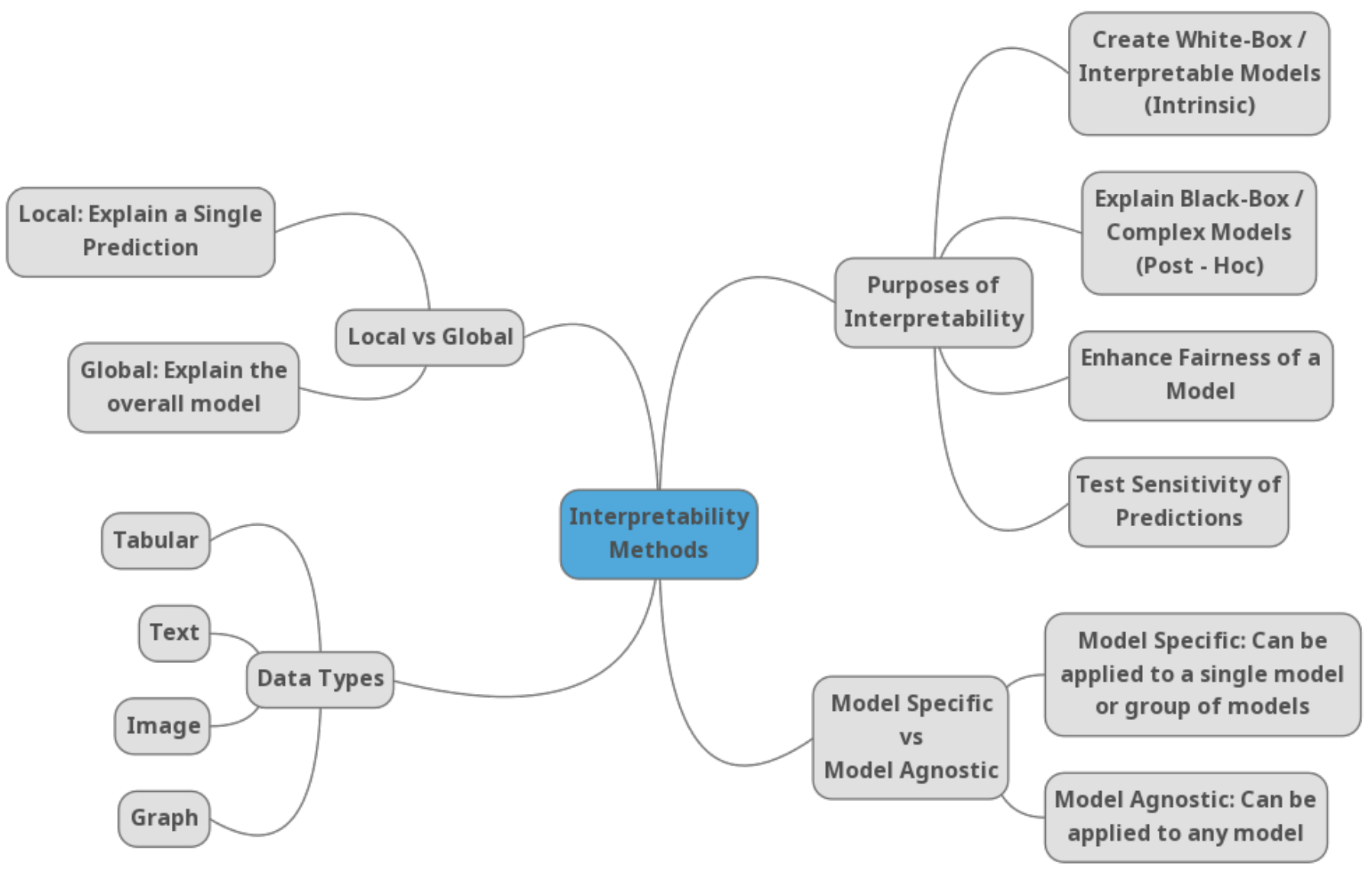}
\caption{Taxonomy mindmap of machine learning interpretability techniques. (image by \cite{linardatos_pantelis_explainable_2021})}
\label{fig:taxonomymindmap}
\end{center}
\end{figure}

When explainability/interpretability techniques are developed it is helpful to be cogniscent of the many aspects of interpretability as illustrated in the taxonomy mindmap in Figure \ref{fig:taxonomymindmap}, page~\pageref{fig:taxonomymindmap}. Machine learning explainability methods may be categorised into four main groups: 

\begin{itemize}[noitemsep]
\item methods for explaining complex black-box models
\item methods for creating white-box models
\item methods that promote fairness and mitigate discrimination
\item methods for analysing the sensitivity of model predictions
\end{itemize}

Explaining black-box models encompasses methods that are known as ``post-hoc interpretability methods'' because they aim to interpret the existing pre-trained models, rather than creating new or modified interpretable models \cite{linardatos_pantelis_explainable_2021}. White-box (or glass-box models), which includes linear regression and decision trees, are limited in performance and adaptability compared to state of the art machine learning models, and hence losing prominence \cite{linardatos_pantelis_explainable_2021}. Significant efforts have been made to address fairness and discrimination, notably through frameworks like Hardt et al.'s equality of opportunity in supervised learning \cite{HardtMoritz2016}, but these methods are not widely integrated into mainstream machine learning practices and are underexplored in non-tabular data contexts. Sensitivity analysis has gained traction following studies on adversarial examples that expose vulnerabilities in deep learning models. Despite the growth in \gls*{xai}, the field still lacks maturity and standardisation, and interpretability methods are not yet a core component of typical machine learning workflows \cite{linardatos_pantelis_explainable_2021}. Moreover, several researchers are calling for the creation of ``transparent'' \gls*{ai} systems where the training data used was serialised on blockchains \cite{baioumy_ai_2024,fernandez-becerra_enhancing_2024,marengo_future_2024} and blockchain-backed marketplaces of \gls*{ai} systems \cite{baioumy_ai_2024,bentley_combination_2024}, indicating substantial potential for future research into explainable AI.

% ----------------------------------------
% Less than Human Performance
% ----------------------------------------
% -------------------------
\subsection{Areas of Less-Than-Human Performance}
% -------------------------
\label{sec:lessthanhumanperformance}

One of the now-classic ways to trick an \gls*{llm} into giving a bad answer, or to show how the technology fails, is the prompt, ``How many times does the letter r occur in the word 'strawberry'?''

Most \glspl*{llm} will answer ``2'', which is incorrect. The correct answer should be ``3''. The \glspl*{llm} get this wrong because they never see the word ``strawberry'' in their input. Instead, they only see a number representing a token for that word. That is, they cannot reason over the word because they only receive a number.

There are ways to work around this problem. For example,
\begin{enumerate}
\item Can you list all letters in the word "strawberry" in the order that they appear?
\item How many times does the letter r appear in the list that you generated?
\end{enumerate}

ChatGPT 4 or 4o will correctly answer ``3''.

New approaches to reasoning in \glspl*{llm} are addressing these limitations while at the same time introducing new performance penalties. ChatGPT o1-preview (intentionally code-named ``Strawberry'') will correctly answer ``3'' to the initial question because it does parse the word and then double check itself. However, as of this writing the process takes around 22 seconds. No doubt additional research will improve that performance.
        
% ------------------------------------
% Research Directions
% ------------------------------------
\section{Research Directions}
\label{researchdirections}
% ----------------------------------------------
% Current Research Topics
% ----------------------------------------------
% ----------------------------------------------
\subsection{Current Research Topics}
% ----------------------------------------------
\label{sec:current}
In this section we present a high level summary of the current topics being researched. The strengths of Web3 and AI may be harnessed to complement weaknesses in the other technology, or they can build on their strengths and synergies to harvest the best of both worlds. We see evidence of both in current research and applications in the areas such as finance, governance, insurance, robotics, security, logistics, transportation and health \cite{hechler_ai_2020, hechler_limitations_2020, elmougy_anomaly_2021, hao-wen_challenges_2023, kumar_artificial_2023, mala_integrating_2023, palaiokrassas_machine_2023, yang_survey_2023, baioumy_ai_2024, xu_vasa-1_2024, yunlong_incentive_2024}.

\subsection{Integration of Web3 \& AI in Business}
% -----------------------------------------------------------------
\label{sec:business}
Kumar et al \cite{kumar_artificial_2023} conducted a bibliometric and content analysis primarily aimed at the integration of blockchain and AI across business functions. They posed four research questions that were addressed in the paper. Three questions related to the identification of what can be considered the most influential papers published to date in various subject areas.The fourth research question sought to identify those business areas that are most promising for the application of Web3 and AI integration \cite{kumar_artificial_2023}.

% -----------------------------------------------------
% Potential future topics / tasks
% -----------------------------------------------------
% --------------------------------------------------------
\subsection{Potential Future Topics \& Tasks}
% --------------------------------------------------------
\label{sec:future}

\begin{figure}[htbp]
\begin{center}
\includegraphics[width=0.5\linewidth]{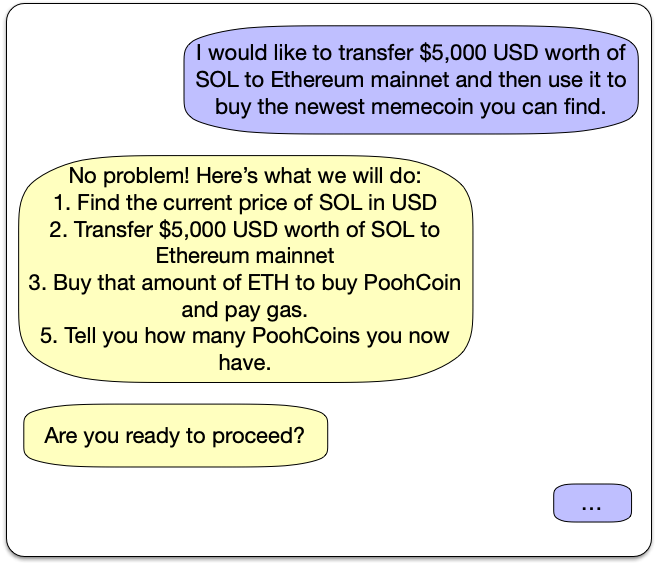}
\caption{Example user chat with a blockchain-aware \gls*{llm}}
\label{fig:userchat}
\end{center}
\end{figure}

A use case of \glspl*{llm} that occurs early to many Web3 developers is to abstract the complexities of currency conversions, transfers, and trades by having an \gls*{ai} automatically generate an execution plan. An example user interaction is illustrated in Figure \ref{fig:userchat}. In this scenario, a user would like \gls*{ai} support to identify the current prices and exchange rates of several currencies, estimate gas costs, develop an execution plan that mirrors human intent, and then execute it upon confirmation. Such a scenario is indeed possible with the obvious caveat that there is much work to be done to create and test the appropriate guard rails. We recommend proceeding very carefully before developing any such economic scenario where a user's wallet is acted upon by \gls*{ai} agents.

It is possible to mitigate the risks associated with moving users' funds by capping the quantities of funds available for transfer via this mechanism and/or by limiting the permissible actions allowed to the agent.

In contrast, we do recommend immediate exploration of \gls*{ai}-assisted user interactions at the wallet level for non-economic interactions (i.e., those activities that do not involve movement of funds) or pseudo-economic interactions (e.g., accessing an oracle for representative price information).

Some useful and immediately-addressable pseudo-economic features include the implementation of \gls*{ai}-driven fraud detection systems, gas fee estimations, and transaction sequencing optimisations.

It is possible that future \gls*{ai}-enabled research on consensus algorithms could result in a breakthrough optimisation, although such a possibility remains just that until the work is done.

\begin{figure}[htbp]
\begin{center}
\includegraphics[width=0.8\linewidth]{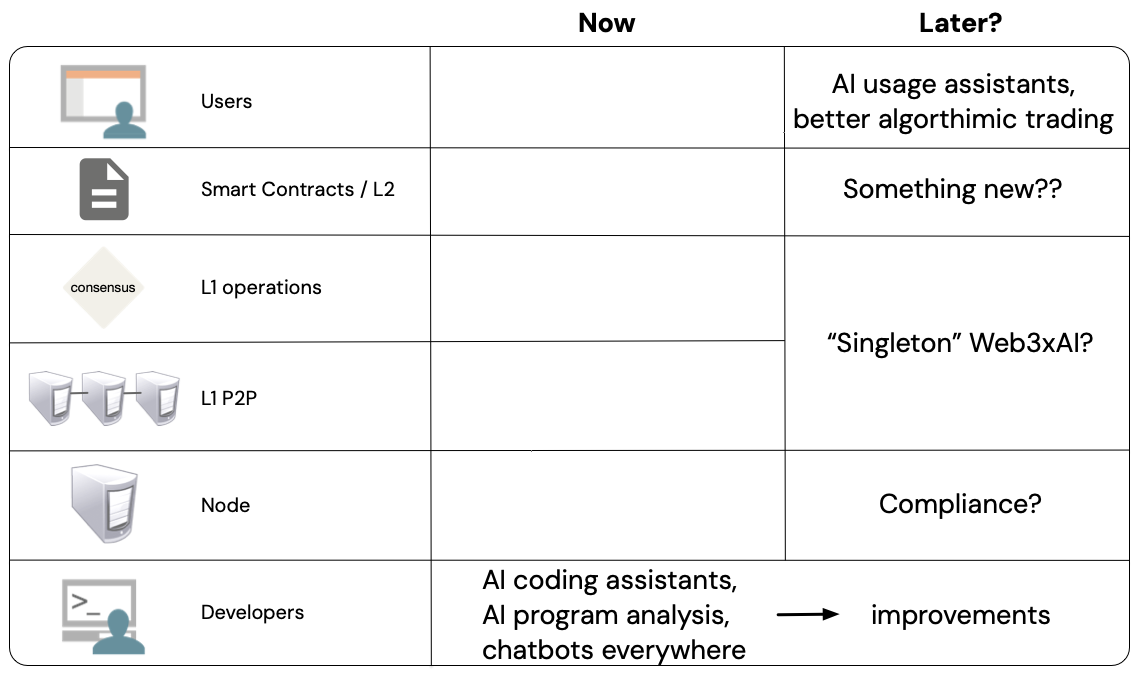}
\caption{Likely applications of \gls*{ai} to the Web3 stack}
\label{fig:possibleapplications}
\end{center}
\end{figure}

Although deployed \gls*{ai} tools are currently most commonly used by software developers in the Web3 space, the research directions above are becoming at least partially clear. We have attempted to summarise the possible near-future applications according to the area of Web3 development in Figure \ref{fig:possibleapplications}.

% ----------------
% Conclusions
% ----------------
\section{Conclusions}
% ----------------
% Conclusions
% ----------------

We have reviewed literature on the \gls*{web3xai} intersection from a variety of sources including academic research, industry research, the open source software development community and industry thought leaders. We have extracted key trends in current \gls*{web3xai} research. We have attempted to describe both the technologies and the possible impacts of their applications in a manner intended to assist decision makers.

\gls*{ai} technology is anticipated to increase labour productivity over the next 10 to 20 years, and to contribute significantly to industrial nations' productivity growth. Importantly to software development organisations, business operations and software development activities make up the majority of anticipated \gls*{llm} impacts on the labour market, which will naturally affect software development organisations disproportionally. It seems that companies related to science, technology, engineering and mathematics (STEM) that embrace the use of \gls*{ai} are likely to outperform those that do not. However, new research has cast doubt on the overall productivity forecasts. Time will tell who is correct.

It has been suggested that \gls*{ai} adoption within corporations should be ``domain focused'' to align adoption with the specific needs within an organisation.

\glspl*{llm} seem to be following the 10/10 rule and not a 1/1 step function, as discussed in the \nameref{sec:introduction}. We most likely will spend the better part of the next decade absorbing and integrating the capabilities brought by \glspl*{llm}.

The nature of work in the information economy is also anticipated to change drastically. We can reasonably expect jobs in the information economy to look very different in coming decades than they do today. For software development companies, this means a transformation in how software is created, tested, and maintained, with \gls*{ai} taking over routine coding tasks, enhancing productivity, and enabling developers to focus on more complex problem-solving and innovative work.

However, any new technology comes with caveats and even warnings. We identified several areas where the current suite of \gls*{ai} technologies do not perform as well as humans, 
%hallucinated in even dangerous ways, 
or failed to live up to their hype. 
%We presented a primer on how \glspl*{llm} work to assist users in working with, and not just relying upon, the tools. 
%We included several examples of both failures and their workarounds. 
%We also suggested new research needed to uncover the full extent of the problem, especially when using \glspl*{llm} as research tools.
We will need to be aware of unintended consequences as they arise. We have suggested a few that we seem likely to experience, but do not pretend to have a perfect vision of the future.

The most immediate applications of \gls*{ai} are disproportionately impacting software developers and testers. They include tools for software creation, test generation, code auditing, program analysis, general knowledge access, drastically improved auto-completion, and general purpose usage assistants.

Using \glspl*{llm} to code, summarise, and suggest new content is rapidly becoming an efficiency argument; individuals can get more done faster. So, what should individuals do with the time saved? Many economists suggest that individuals should spend their time doing more of whatever their job is \cite{furman_ai_2019}. We suggest that when technology is changing really rapidly (as it is now), individuals will need to spend a lot more time adjusting to those changes. That might include learning new languages, \glspl*{api}, tooling, techniques, methodologies, etc. So, we suggest the saved time should be used right now in navigating the changes caused by \glspl*{llm} and other \gls*{ai} integration. That time investment seems most likely to prepare those organisations that do so for the next round of changes to come.

Applications of \gls*{web3xai} exist at all layers of the Web3 stack. Market opportunities are clearly a subset of the possible application areas.

The application areas with the most immediate utility remain in text and code generation, text summarisation, and code auditing. These areas are anticipated to continue to improve rapidly so keeping up with those changes should remain a priority for the foreseeable future.

Opportunities for the application of \gls*{ai} into Web3 systems identified by this work include:
\begin{itemize}

\item Short term
\begin{itemize}
\item \gls*{ai}-driven features at wallets and \glspl*{api} to improve user interfaces, customer support and user experience
\item \gls*{ai}-driven auditing enhancements for Web3-specific components such as smart contracts
\end{itemize}

\item Medium term
\begin{itemize}
\item \gls*{ai}-based fraud detection systems
\item the use of \gls*{ai} for gas fee estimation and transaction sequencing optimisations
\item multi-jurisdictional legal compliance assistance at the \gls*{L1} client level
\end{itemize}

\item Long term
\begin{itemize}
\item ``singleton'' \gls*{web3xai} \glspl*{L1} and \glspl*{L2}
\end{itemize}

\end{itemize}

% --------------------------
% Acknowledgements
% --------------------------
\section{Acknowledgements}
We thank Consensys Software Inc for funding this research.   

% Bibliography
% -----------------------------
%\clearpage                               % Needed to avoid tables showing inside References section
%\nocite{*}
\bibliographystyle{eptcs}
\bibliography{aireferences}

\end{document}